\documentstyle[psfig,conf_iap,]{article}
\begin{document}
\heading{%
%
Probing the OVI forest
%
} 
\par\medskip\noindent
\author{%
Greg L. Bryan$^{1}$, Taotao Fang$^{2}$
}
\address{
Oxford University, Astrophysics, Keble Road, Oxford, OX1 3RH, UK.
}
\address{
Center for Space Research, MIT, Cambridge, MA, 02139, USA.
}

\begin{abstract}
  Recent FUSE and STIS observations of O~{\sc vi} absorption at low
  redshift along the sightline to distant quasars have been
  interpreted as the signature of a warm-hot diffuse component of the
  intergalactic medium (IGM).  In these proceedings we show that the
  predicted numbers of such absorbers in numerical simulations agree
  with the observed characteristics, lending support to this idea.  We
  find that collisionally ionized lines tend to be stronger and wider,
  while photo-ionized absorbers are weaker and narrower.  We also find
  that a comparison of the predicted distribution of line widths to
  that observed is marginally too low.

\end{abstract}

\section{Introduction}

The Ly$\alpha$ transition of five-times ionized Oxygen is an important
tracer of diffuse matter in the universe, both at high redshift where
the transition falls in the optical and now --- thanks to new
high-resolution spaced-based UV-spectroscopy --- at low and moderate
redshift as well.  Absorption lines due to intervening gas in
high-redshift quasars are an indispensable way to detect low density
material because the signal is proportional to the integrated column
density of the element.  In contrast, X-ray emission grows as the
second power of the local density.  This means that the low-density,
hot filaments predicted in numerical simulations \cite{co99,dave} are
much more accessible to absorption line studies than to direct X-ray
detection.

\section{Predicting the number of O~{\sc vi} lines}

In these proceedings, we use numerical simulations of a
cosmological-constant dominated universe ($\Omega_\Lambda=0.7,
\Omega_b = 0.04, h=0.67$) to make predictions of O~{\sc vi} absorption
line statistics.  We use an adaptive mesh refinement code
\cite{bryan99} to model a 20 $h^{-1}$ Mpc box with up to 10 kpc
resolution (in small regions).  A drawback of this simulation is that
while it includes hydrodynamic shock-heating, it does not model star
formation and stellar feedback.  Therefore, it does not
self-consistently pollute the IGM with metals.  In order to make
predictions for the O~{\sc vi} distribution, we must assume a
distribution of metals.  We do this by parameterizing the results of
\cite{co99}, who found in their simulations a strong correlation
between local density and metal abundance.  We have also experimented
with a spatially constant metal fraction of 10\%, which does not
substantially change the results described below.

\begin{figure}
\centerline{\vbox{
\psfig{figure=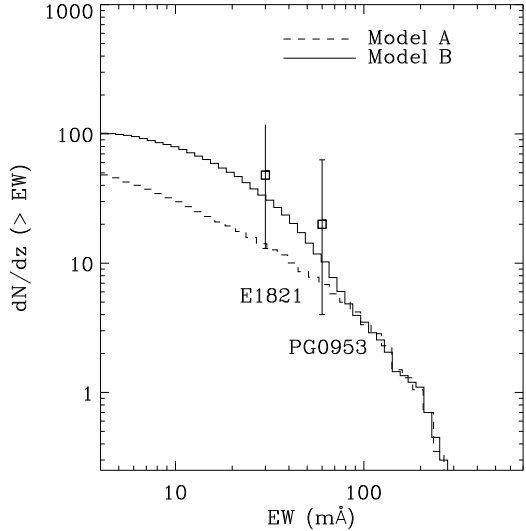,height=7.cm}
}}
\caption[]{The predicted cumulative number of O~{\sc vi} absorption
  lines.  The solid line (Model B) includes both photo- and
  collisional ionization, while the dashed line (Model A) is computed
  assuming only collisional ionization.  Two observation points are
  plotted with 1$\sigma$ error bars from \cite{ts2000, tsj2001}.
}
\label{fig:dndz}
\end{figure}

In Figure~\ref{fig:dndz}, we plot the predicted number of O~{\sc vi}
absorption lines as a function of equivalent width.  In order to
investigate the relative effects of collisional and photoionization,
we plot the results with and without a uniform ionizing background (as
predicted in \cite{hm96}).  The collisional ionization-only model
marginally matches the observations, while the combined version
provides excellent agreement with observations.  In either case, lines
with equivalent widths larger than about 70 m$\AA$ arise only in
collisionally ionized gas.

\section{The distribution of line widths}

\begin{figure}
\centerline{\vbox{
\psfig{figure=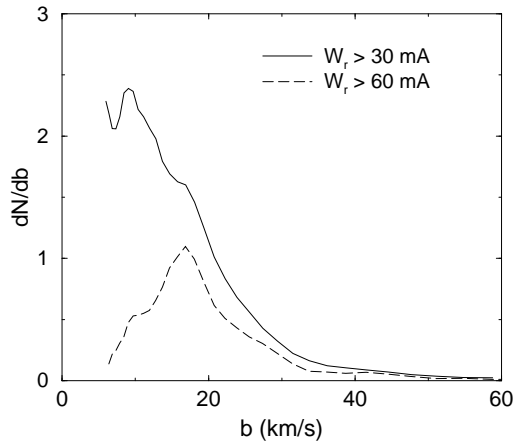,height=6.cm}
}}
\caption[]{The predicted distribution of Doppler line widths for two
  different minimum cutoffs in the line strength.}
\label{fig:dndb}
\end{figure}

If the raw number of absorption lines agree with observations,
what about the distribution of their widths?  In
Figure~\ref{fig:dndb}, we plot the prediction for this distribution,
assuming minimum cutoffs of 30 m$\AA$ and 60 m$\AA$.
Clearly these two distributions are quite different, with the higher
cutoff curve coming mostly from collisionally ionized regions, while
the lower cutoff includes contributions from both ionization
mechanisms.  This plot confirms previous speculation (\cite{ctoj, fb01})
indicating that collisionally ionized lines are hotter and hence have
larger Doppler parameters than photo-ionized regions.

While it is difficult to compare this to observations due to the
differing resolutions and analysis techniques, a recent compilation of
low-redshift O~{\sc vi} lines provides a useful point of comparison
\cite{tripp01}.  This sample, which includes a diverse set of systems,
has a median $b$ parameter of 22 km/s.  The two distributions in
Figure~\ref{fig:dndb} have medians of 13 km/s (for the 30 m$\AA$ cutoff)
and 18 km/s (for 60 m$\AA$).

\section{Discussion}

There is a reasonably strong case that the O~{\sc vi} absorption lines
are tracing the filaments of hot gas long predicted by numerical
simulations of structure formation.  Here we show --- in agreement
with a recent preprint \cite{ctoj} --- that the predicted number of
absorbers roughly agrees with the number observed.  

Most of the absorbers are found in systems with densities typically
5-100 times the mean density (see \cite{fb01} for a more complete
discussion of this point).  These are not primarily galactic systems,
but are the web of filaments that link collapsed objects.

While the total numbers are in good agreement with
observations, the line widths (as measured by the Doppler $b$
distribution) are somewhat low.  Although it is not clear what the
appropriate lower cutoff is for the observational sample discussed
above, it seems likely that the simulations predict median $b$
parameters that are lower than that seen in FUSE and STIS samples.

If this disagreement persists, there are a number of possible
explanations.  The first is that the numerics are incorrect.
Certainly the filaments are marginally resolved, and it is
possible that improved simulations will refine the predicted value (as
occurred for the Ly$\alpha$ forest results at high redshift).  The
second is that energy ejected from galactic systems increases the mean
temperature of the IGM, which would boost the $b$
parameters in Figure~\ref{fig:dndb}.

\begin{figure}
\centerline{\vbox{
\psfig{figure=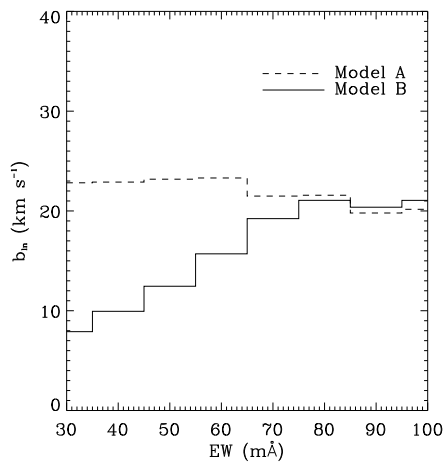,height=6.cm}
}}
\caption[]{The peak of the Doppler $b$ distribution as a function of
  the equivalent width cutoff used in the sample.  The models are as
  before (A is collisional-only, while B includes photo-ionization).}
\label{fig:bln}
\end{figure}

A final possibility is that this is telling us something about the
strength of either the ionizing background or the value of the mean
baryon density (or the metallicity).  Varying either of these could
decrease the number of photo-ionized systems which make it above the
30 or 60 m$\AA$ cutoff and so increase the proportion of lines which
are collisionally ionized.  Since these hotter systems have higher $b$
values in general, this would boost the median $b$ value.  In
Figure~\ref{fig:bln}, we have used various line-strength cutoffs and
then fit the resulting $b$ distribution to a log-normal profile.  For
the model which includes photo-ionization, increasing the cutoff leads
to an increase in $b_{ln}$ (which is nearly equal to the median).  It
is clear that the distribution of line widths (and absorption line
studies in general) will tell us important information about the
physical condition of the IGM.

\acknowledgements{This work is supported
  in part by contracts NAS 8-38249 and SAO SV1-61010, and NASA through
  Hubble Fellowship grant HF-01104.01-98A from STScI.  Simulations
  were carried out on the Origin2000 at NCSA.}

\vfill

\begin{iapbib}{99}{
\bibitem{bryan99} Bryan, G.L. 1999, Comput. in Science and
  Engineering 1:2, 80
\bibitem{co99} Cen, R. \& Ostriker, J.P. 1999, \apj 514, 1
\bibitem{ctoj} Cen, R., Tripp, T.M., Ostriker, J.P., Jenkins,
  E.B. 2001, astro-ph/0106204
\bibitem{dave} Dav\'e, R., Hernquist, L., Katz, N., Weinberg,
  D.H. 1999, \apj 511, 521
\bibitem{fb01} Fang, T. \& Bryan, G.L. 2001, \apj Letters, in press (astro-ph/0109276)
\bibitem{hm96} Haardt, F. \& Madau, P. 1996, \apj, 461, 20
\bibitem{ts2000} Tripp, T.M. \& Savage, B.D. 2000, \apj, 542, 42
\bibitem{tsj2001} Tripp, T.M., Savage, B.D., Jenkins, E.B. 2000, \apj,
  534, L1
\bibitem{tripp01} Tripp, T.M. 2001, to appear in ``Extragalactic Gas at Low
  Redshift'', ASO Conference Series, eds. J.S. Mulchaey \& J. Stocke (astro-ph/0108278)
}
\end{iapbib}
\end{document}